# Tracing Complexity in Food Blogging Entries

Maija Kāle[1[0000-0002-6951-9009 ]] and Ebenezer Agbozo[2[0000-0002-2413-3815]]

[1] University of Latvia, Riga, Latvia
[2] Ural Federal University, Yekaterinburg, Russian Federation
maijakale@gmail.com

**Abstract.** Within this paper, we focus on the concept of complexity and how it is represented in food blogging entries on Twitter. We turn specific attention to complexity capture when looking at healthy foods, focusing on food blogging entries that entail the notions of health/healthiness/healthy. We do so because we consider that complexity manifests hedonism - that is the irrational determinant of food choice above rational considerations of nutrition and healthiness. Using text as a platform for our analysis, we derive bigrams and topic models that illustrate the frequencies of words and bi-grams, thus, pointing our attention to current discourse in food blogging entries on Twitter. The results show that, contrary to complexity, that the dominating characteristics in healthy food domain are easiness and speed of preparation, however, rational and health related considerations may not always take precedence when the choice is determined. Food blogging entries show surprisingly little account of healthy food as being tasty and enjoyable. With this we aim to contribute to the knowledge of how to shape more healthy consumer behaviors. Having discovered the scarcity of hedonic connotations, this work invites for further research in text-based information about food.

## 1 Introduction

The changing nature of humans from hunter-gatherers to super-consumers [15] has its implications. Lifestyle-induced obesity, Type 2 diabetes and cardiovascular diseases are on the rise, leaving public health policy makers with little success. Most of the debate within the food sector has been a chicken-versus-egg type of discussion on who is to blame first – the consumer that prioritizes sweet and fatty fast food or the multinational corporations, i.e. the big food members that race for providing increasingly sugary, salty and fat-intense foods.

Food consumption is a complex phenomenon involving many aspects that are easier or harder to be captured. All in all, food choice is highly contextual: accessibility, cultural background and habits all play heavier or lighter roles in food-related decisions. Besides that, the weights of these determinants are unpredictable and can change from case to case. Furthermore, when it comes to a consumer's choice of food, the control over their decisions is limited. "Although eating is an action of which the individual may be aware, control of eating habits is not necessarily explicit. In contrast, consumers frequently choose their food products in an unconscious manner" [9].

With our research we are focusing on the drivers behind the consumers' choice of food, stepping out of rational thinking domains related to it. We focus on hedonism as an irrational driver for the consumers' choice and explain it by the level of complexity that determines the likeability of the food that results in its choice as described in the forthcoming paragraphs.

### 1.1 The Formula Behind the Food Choice

One of the guiding principles to think about the mechanisms behind the food choice is through the prism of hedonism. Food intake is a hedonistic act which is hard to predict. Nevertheless, by now it is clear that hedonism dominates other types of considerations – such as healthy food choice or moderate size of portions. "A growing body of research has shown that our hedonic responses to foods and beverages are not simply determined by their physical and chemical properties [..] To fully understand consumers' food behaviors, it is clearly important to investigate the factors that affect the hedonic responses to foods and beverages" [12].

Looking at digital media we can see that food-related hedonism or certain aspects of it have been expressed and documented in a growing amount of textual and visual information: the increasing number of food images, food blogs, the rise of celebrity chef justifies the notion that "We eat first with our eyes" (Apicius, 25 BC) [15]. This is explained by the fact that the "human brain evolved during a period when food was much scarcer than it is now" and consequently "our brains learnt to enjoy seeing food, since it would likely precede consumption" [15].

However, the modern consumer seeks for a more elaborate sensorial experience than purely seeing the food. Instead, the food we see has to be beautiful, pleasant and within the right ambience. This quest is well documented in the growing multitude of food images on Instagram, the art of plating and multisensorial marketing, which introduce such new modus operandi as expanding the field of knowledge in cross-modal association research [19].

The modern-day hedonism when it comes to food is way wider than just the pleasure of seeing the food as such. It is elaborated, multisensorial, and frequently balances on the edge between the real and virtual worlds [11, 15].

The importance of hedonism and deriving pleasure from food consumption has also been confirmed in the cases of designing food experiences for space tourists [11], encouraging "researchers to think about the positive relationship between humans and food that should be augmented through technology, rather than just focusing on corrective technology that focuses only on the negative practices in humans' relationship with food" [11]. This suggestion is of a pivotal importance – while there is a growing multitude of food data, be it images or descriptive texts, we do not have an extensive body of knowledge that would reveal the formula behind the positive relationship between humans and food. At this moment, the rationale behind the food-related decisions is still uncaptured, and inefficiency that exists is better documented via corrective aspects of human and food relationships – such as unhealthy diets and food waste.

## 1.2 Complexity Explained

One of the concepts applied in discussing hedonic experiences in food consumption is the concept of complexity, the definition of which is rather intuitive. "Complexity is commonly talked about as a desirable attribute of the consumer's experience of food and drink" [16]. Potentially the best outplayed complexity has been within the world of wine tasting where "the single greatest standard used in assessing the quality of a wine is complexity. The more times you can return to a glass of wine and find something different in it – in the bouquet, in the taste – the more complex the wine. The very greatest wines are not so much overpowering as they are seemingly limitless." [16].

Another concept, which is more widely used to describe certain aspects of a joyful/hedonistic experience has been the concept of "flow" [4], which defines such an immersion within an activity that the sense of time gets altered considerably. The concept of "flow" could be highly interesting when discussing the intersection of food consumption and the sense of time, however, in our work we are more interested in exploring the preceding step – namely, the very point of choosing the food rather than the moment of its consumption. Moreover, the concept of "unity" somewhat resembles the concept of "flow", providing an instrumental tool that explains how "signals from the different senses are encoded into a unified object/event representation" [2].

While the concepts of "flow" and "unity" rather focus on different senses turning into one monolith experience, the concept of "complexity" allows for capturing the diversity of the given experience.

We choose complexity as an explanatory variable of hedonism as the major concept for analytics due to its multiple layers of definition, as well as its compatibility with other concepts (such as healthiness, which will be explained in the next sub-chapter). First, looking at the concept of complexity and its definition, we can see that "complexity can be operationalized in ingredients, in preparation, or in the flavor experience" [16]. In other words, we can talk about complexity of food from the perspective of it being prepared:

- ingredients (the number of ingredients, where more ingredients mean more complex food);
- preparation (the time spent for preparing the food, where a longer preparation time means more complex food);
- consumption, which results in flavor experience (the greater the number and the variety of flavor descriptions, the more complex the food).

Such classification of complexity allows us to build further layers of "complexity in chemical senses, perceived complexity and the ability of a flavor experience derived from our exposure to the chemical senses to evoke 'cerebral (cognitive) and sensorial responses' that is what leads us to consider it as 'complex' rather than 'simple'" [17]. Furthermore, we could even look into "inferred complexity rather than necessarily perceived complexity, given the differing perceptual attribute (not to mention inferential processing) that would seem to underlie such judgements" [17].

To conclude, the concept of complexity invites us to shape it in accordance with our needs and carve out the exact meaning that would serve our analysis best. For the time being we will rely on complexity described through ingredients, preparation and expression of flavor and their definitions as described above accordingly.

### 1.3 The Battle Between the Healthy and the Tasty

While complexity could be the leading prism through which to observe the level of hedonism, another important divide is a dichotomy between healthy and unhealthy food. One of the key hindrances towards consuming healthy food is the unhealthy=tasty intuition [9]. “Previous attempts on the part of policy makers to induce healthy food choices by educating consumers have not yet led to major improvements or fundamental changes in actual behavior [..] the objectives of health and taste often conflict - and that taste usually prevails in food decision making.” [9]

This intrinsically implies that the level of hedonism is higher in unhealthy foods, which are considered tastier than the healthy ones. Choosing chocolate as one of the foods best depicting the level of hedonism, we can see that “a ‘reduced fat’ label, for example, negatively affects the acceptance of chocolate bars” [9]. This gives additional motivation to look into complexity as a driver for food choice and points to the potential of assessing the level of complexity in healthy foods.

Imbuing healthy food with the desired portion of hedonism and presenting it as a solution to unhealthy diets has its grounding in the multisensory field. “Research on overeating assumes that pleasure must be sacrificed for the sake of good health. Contrary to this view, the authors show that focusing on sensory pleasure can make people happier and willing to spend more for less food, a triple win for public health, consumers, and companies alike” [3]. Thus, pleasure and hedonism matter, and there is potential for embedding and highlighting those qualities in healthy food.

### 1.4 Food Blogging Entries: Words and the Company They Keep

Going back to 25 BC and the idea that “we eat first with our eyes” (Apicius, 25 BC) [15] it could be reformulated into “we first eat with our mind”. Here we mean that a modern human is consuming texts and stories *about* the food as eagerly as images *of* it. Recent studies have indicated to what extent textual information or storytelling about food influences the multi-sensory experience of food intake. “In particular, according to one analysis of online restaurant menu descriptions, the average price of a dish in the US was found to go up by 6 cents for each additional letter in the dish description” [16]. The concept of words that count is also well documented in the growing number of food blogging and food-related texts in social media. Taking into account the growing popularity of food blogging, we base our analysis on textual information about food, which allows us to conclude the way food is being depicted in those texts.

We have previously analyzed recipes of various cuisines, looking at complexity from the point of view of the listed ingredients. While recipes mainly consist of the lists of ingredients, the analysis of food blogging entries could potentially provide us with more elaborated information. We aim to dig into stories/texts available on social media and appraise to what extent the quest for hedonism and complexity of experiences is reflected in the healthy food domain judging from the ingredients, the preparation and the flavor experiences described in the texts.

## 2 Research Goals and Methodology

With this research we aim to contribute to the understanding of hedonism and the positive relationship between humans and food. Therefore, first we will try to look into the components of hedonism and see what defines an experience as hedonic when it comes to humans and food. We look into complexity as one of the explanatory factors for hedonism and contrast it with the aspect of healthiness. The object of our analysis is the available textual data on food, and the research question we aim to answer is: **how can we utilize the data available in order to move towards more healthy diets and, consequently, to a higher life quality**?

With such analysis, our study aims at testing the hypothesis that healthy foods are depicted as less complex, and, thus, less hedonic. We argue that healthy foods lack the notion of complexity, thereby becoming inferior to less healthy alternatives that accommodate the notion of complexity. Furthermore, we consider that by increasing the level of complexity, the level of hedonism as associated with healthy foods could rise, thus changing the landscape of consumer's choice towards a healthier one.

To answer our pre-set research goals, we employed the technique of text mining for knowledge extraction. As reiterated by Khatai et al. [7], the IBM Knowledge Center describes text mining as "a process of analyzing data to capture key concepts and themes, and uncover hidden relationships and trends without prior knowledge of the precise words or terms that authors have used to express those concepts". The technique has been employed in a countless number of studies, such as evaluating customer satisfaction with hotels, predicting hospital admissions based on emergency department medical records, and also understanding customer satisfaction at large [8, 14, 21].

Using the Twitter API (application programming interface) in the R programming language, keyword searches were made for posts in English language on the following topics: '#healthyfood' and '#recipeoftheday'. Taking into account the limitations on data request quota, 28 509 observations were obtained consisting of posts made between late December 2019 and early January 2020. Twitter was selected due to its availability and the open nature of its information. Integrating the Twitter API into the text analysis research has been recommended in numerous studies dealing with sentiment analysis, topic modelling, or knowledge extraction [1, 5, 22].

Figure 1 illustrates the research process, from data collection to visualization. After performing data cleaning techniques such as "stop word" removal as well as using regular expressions to eliminate punctuations, profile names and URLS, a total of 23 950 observations were obtained.

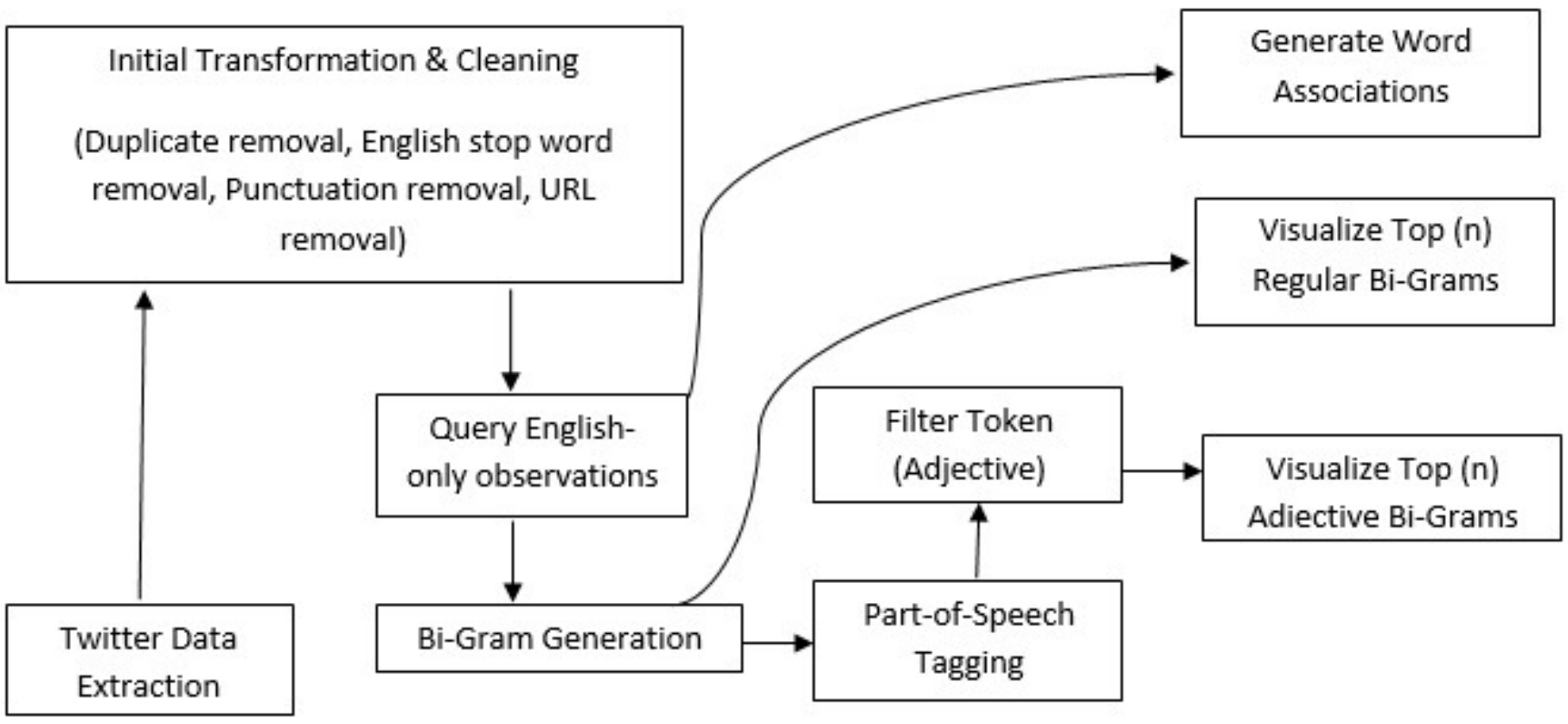

**Fig. 1.** Research methodology and knowledge extraction procedure.

To extract meaningful contextual information from our dataset, we employed n-grams, or, to be more precise, bigrams. N-grams are a sequence of N characters [6]. Despite the possibility of obtaining a great amount of irrelevant observations, n-grams have been noted to be advantageous over isolated words due to the fact that they are more informative and practical [20]. Another method for generating bigrams in our study was also extracting adjectives and their subsequent terms. Frequencies were generated, followed by a visualization.

The final text mining technique employed in our study is word associations. This technique has been used by linguistic and poetry researchers to automatically compose poetry based on a given document as inspiration [18] as well as in bridging the syntactic and lexical chasms in question answering systems [13]. In our case we sought to find out associations of words such as: (1) “healthyfood” and “healthy” supported by terms such as “sustainable”, “nutritious”, and “tasty”; (2) “vegan” and “vegetarian” supported by terms such as “lowfat”, “healthy” and “tasty”.

When referring to blog entries on Twitter within the context of this research, we are referring to one-tweet-long micro-blog entries. This aspect is relevant as longer texts that exceed the standard length of a tweet might lead to different results.

## 3 Results

First, we will present the bi-gram results and then we will look into the details of word associations.

As can be seen in Figure 2, which reflects bigrams with adjectives-first word combinations, the key messages that are shared when talking about healthy food are the following:

- aspects of easiness / easy to cook/ easy to follow the recipe are emphasized
- time-saving aspect in food preparation is emphasized
- tastiness is expressed through such words as “delicious” and “favorite”
- a seasonality aspect is apparent

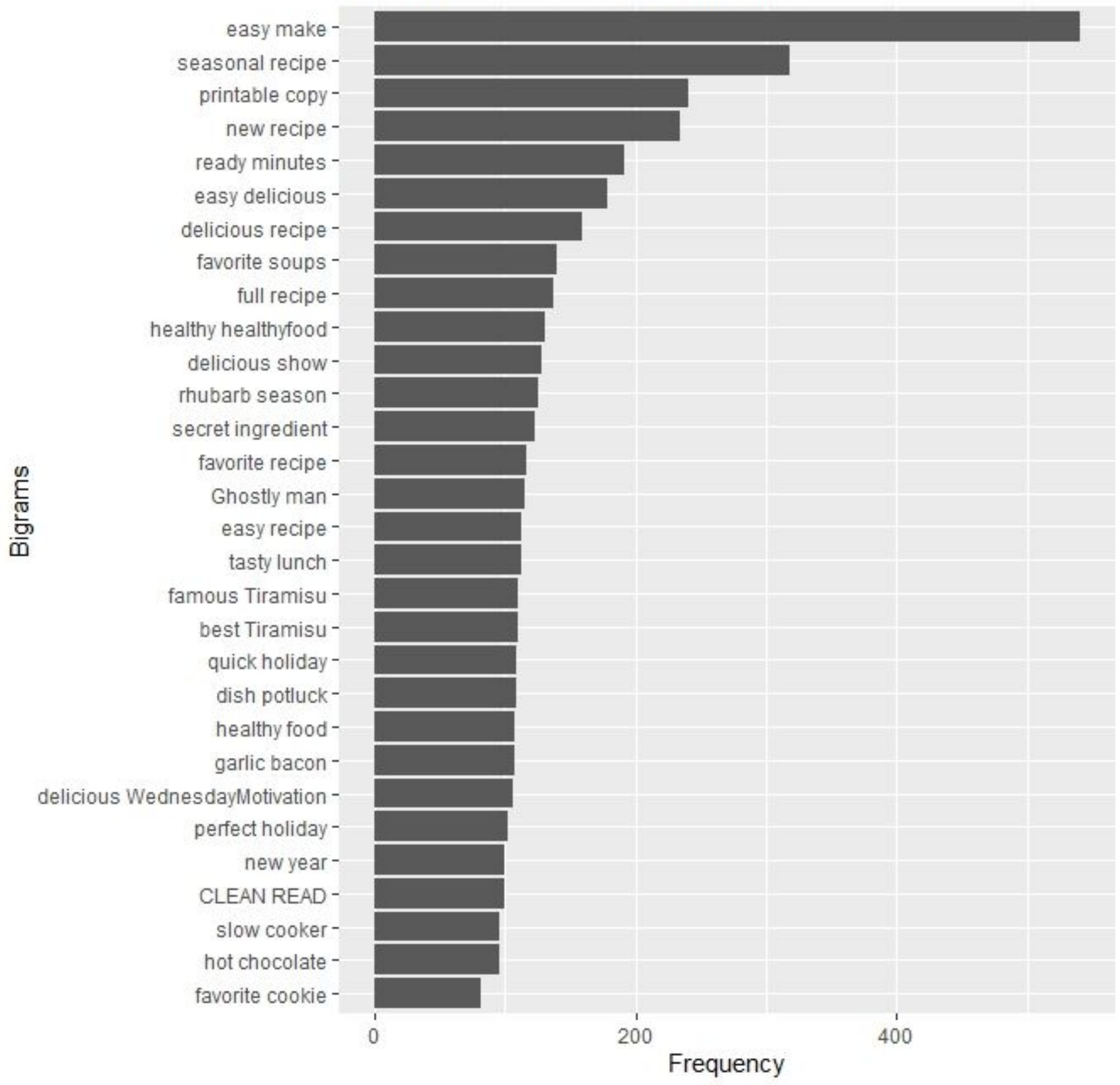

**Fig. 2.** Frequency distribution of Top 30 bigrams with adjectives and regular terms.

In sum, Figure 2 signifies the domination of food preparation aspects – the ease of cooking, the easily followable recipes, and the time used – over all other aspects when it comes to healthy food

The expected flavor experience is documented in a rather laconic manner – just as a promise of delicious food, without any other expressions denoting taste or flavor. To some extent, bigrams containing the words “favorite” or “famous” could be considered indicative of the hedonic realm, but even so the expression could be best described as laconic.

The seasonality aspect that appears in some of the bigrams could be considered related to health, and further still – to planetary health (through consuming locally available foods and saving CO2), potentially shaping some flavor expectations as well.

As concerns the term “Ghostly man” in the food-related bigrams, this reference which leads onto a “true short story” with an embedded recipe[1] testifies to the thematic breadth of the food blogging, which exceeds the constraints of purely food related matters. In this case, literature serves as an additional trigger and an intriguing factor.

1 https://twitter.com/JessieBTyson/status/1204957534039105536/photo/1

Similar trends are also reflected in Figure 3 (bigrams without necessarily adjectives-first word combinations). It contains additional hints inviting to click on the embedded links and proceed onto the cooking videos.

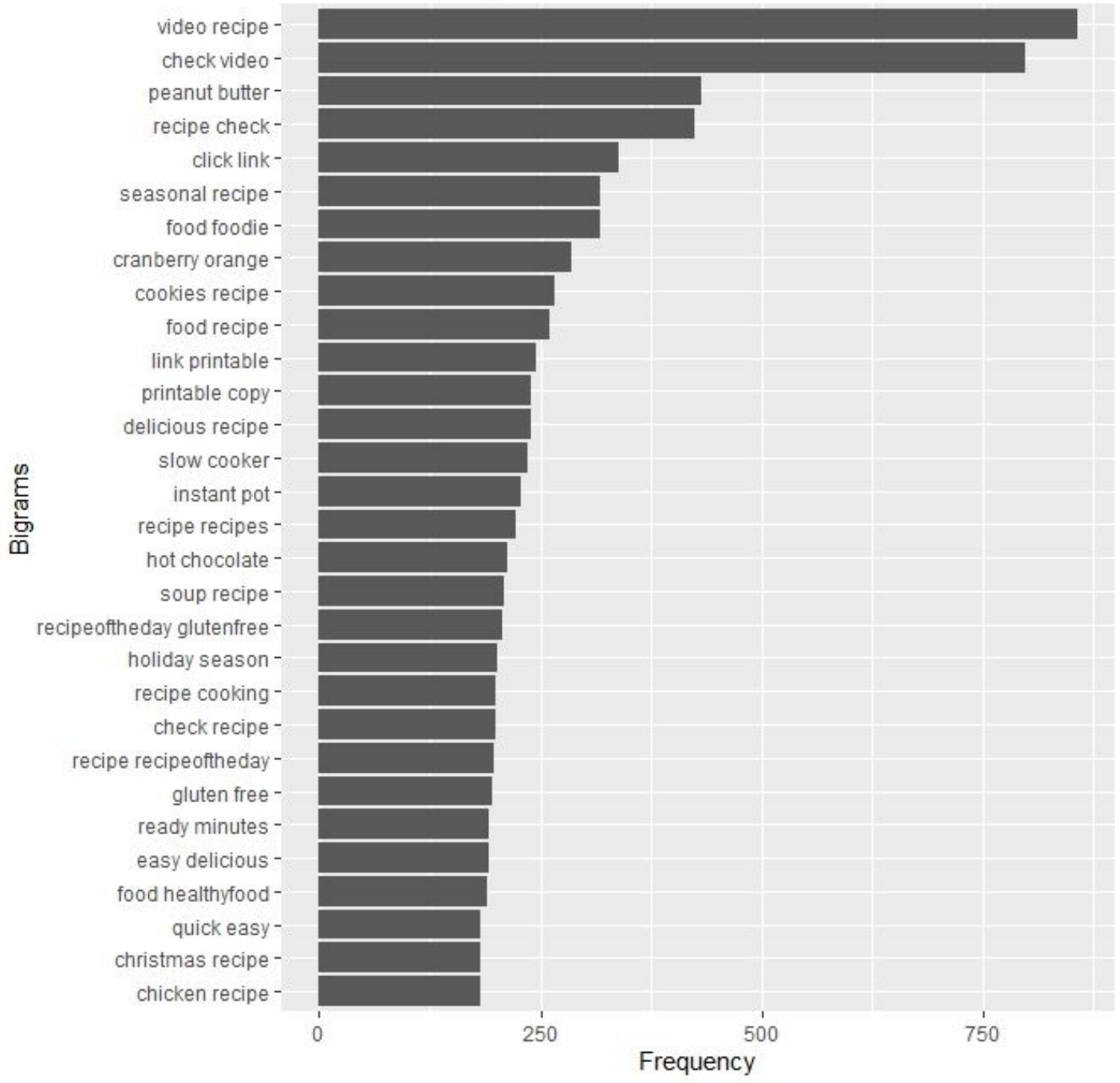

**Fig. 3.** Frequency of Top 30 regular term bigrams.

It would be relevant to explore to what extend hedonism is reflected in videos and images of healthy food but this we leave to further research.

In order to explore deeper the context in which “healthyfood” and “healthy” appeared when supported by terms such as “sustainable”, “nutritious”, and “tasty”, let us look at the Graph A.

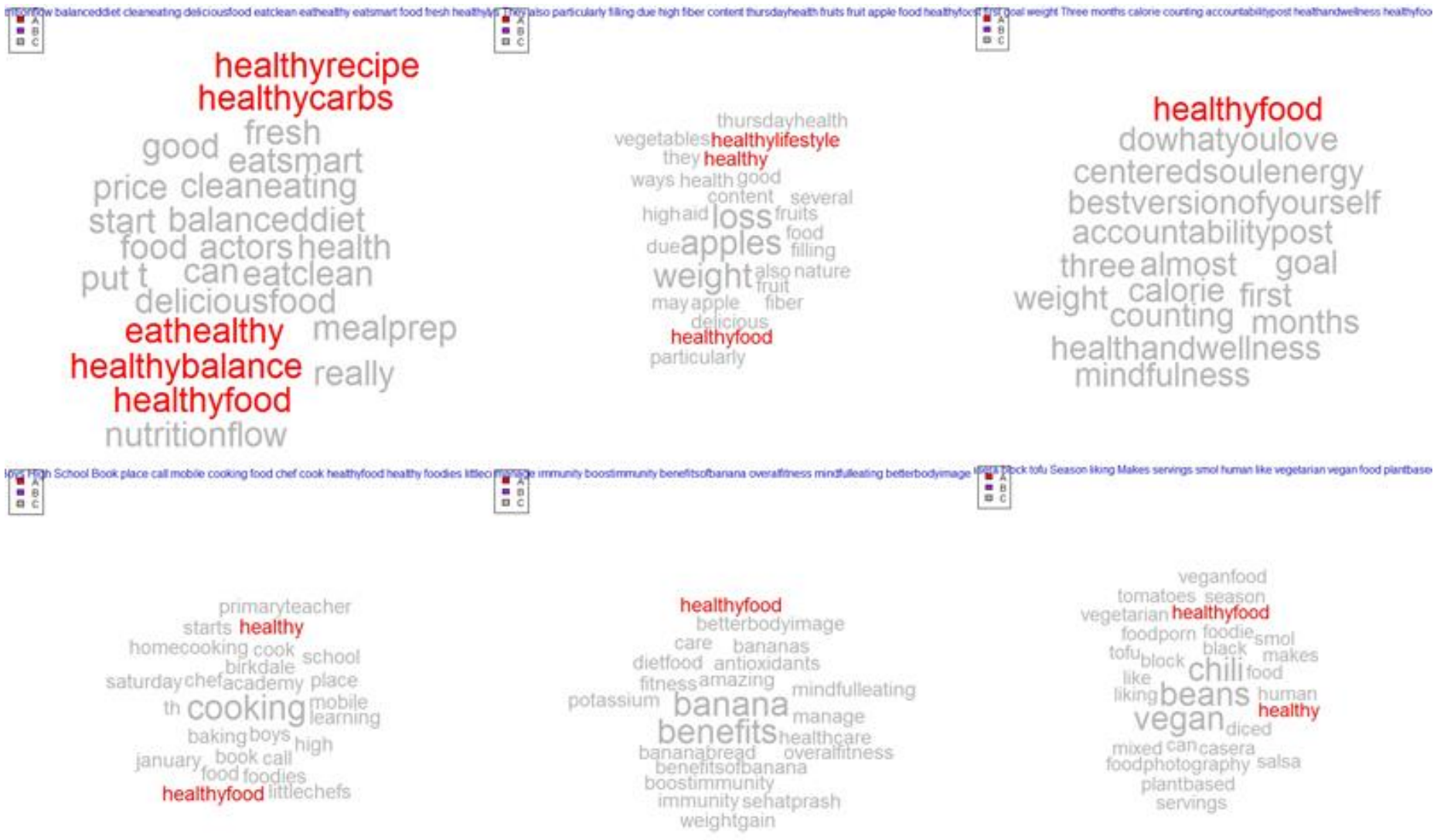

**Fig. 4.** Word associations cloud – healthy, healthy food, and nutritious.

As it transpires in Figure 4, the terms related to health and healthy are more related to nutrition considerations and weight loss, including such terms as “potassium”, “fiber”, “goals”, “calories”. The only term signifying some aspect of hedonism is “delicious”, which appears rather rarely.

For experimental reasons we were also looking how the words “vegan” and “vegetarian” were supported by terms such as “lowfat”, “healthy” and “tasty”. Graph B shows the results.

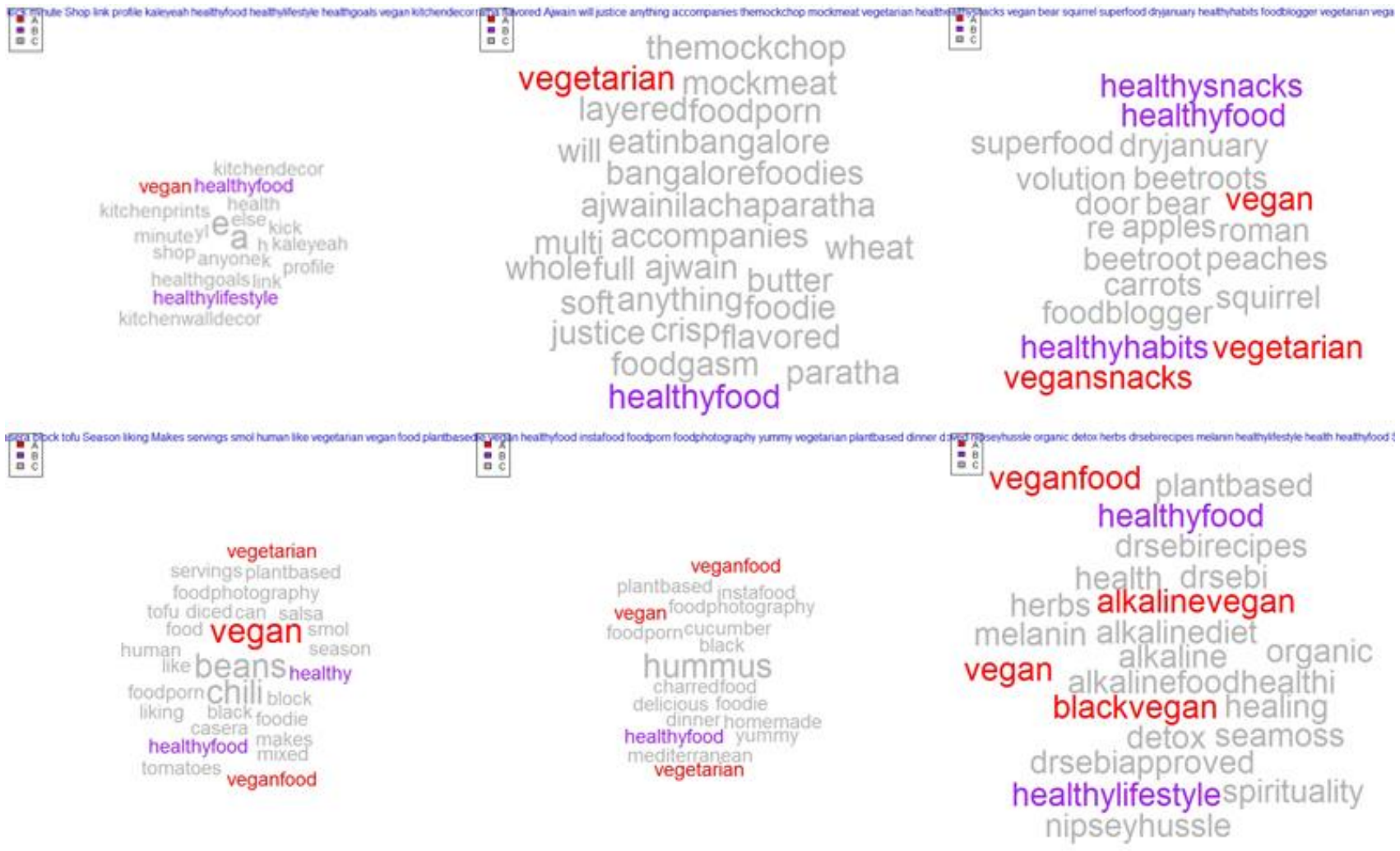

**Fig. 5.** Word associations cloud – vegan, vegetarian, low fat, healthy, and tasty.

As can be seen from Figure 5, neither did the query with "vegan", "lowfat", "healthy" and "tasty food" generate rich representation of terms related to hedonism. Instead, it yielded terms related to rational choice of food – such as "healthy choice", "alkaline diet" etc. The closest term to hedonism in this case appears to be "snacks" – something to treat oneself with while on an alkaline or a vegan diet.

Summing up all the results, we can see that healthy food is starkly related to rational choices – such as nutritional considerations, calories, weight issues, also the ease of cooking. Associative links with hedonism are sparse, being expressed only through a few words – "delicious", "favorite", "famous", "snacks". This leads to the conclusion that when talking about healthy foods, the emphasis is put on rational rather than hedonic aspects, and those rational aspects are accordingly embedded in the texts authored by food bloggers.

## 4 Conclusions

With this research, we aimed to prepare the basis for enhancing consumption of healthy foods. Having made an analysis of the level of complexity of texts describing food, we have ascertained the lack of hedonism-denoting terms in healthy food descriptions.

We did this by looking at the complexity as a necessary precondition for a hedonic food experience, viewing the complexity of food from the point of view of its ingredients, preparation and flavor experience, the latter of which potentially expresses the hard-to-define concept of complexity most fully.

By using the technique of word associations, we first did a bigram analysis and then examined two word-association extractions in order to determine what the word combination "healthy food" is usually accompanied with. It turns out that this word combination tends to go together with the words denoting rational choice and very few words that would signify pure pleasure of consuming the food.

With this we conclude that the current food blog entries of Twitter tweet length related to healthy food do not focus on taste aspects and contain few references to hedonic expressions, focusing rather on "simple and easy" and less so, or almost not at all, on "complex and enjoyable".

Lastly, this research also contributes to the discussion on the extent to which concepts that are used in cognitive science domain are operational in the quantitative analysis of texts.